\documentstyle[prl,twocolumn,aps]{revtex}

\begin{document}

\twocolumn[\hsize\textwidth\columnwidth\hsize\csname @twocolumnfalse\endcsname

\draft

\title{ {\bf Transmission resonances 
on metallic gratings with very narrow slits}   }

\author{ 
 J.A. Porto$^1$, F.J. Garc\'{\i}a-Vidal$^2$, 
 and J.B. Pendry$^1$ \\ } 

\address{  
   $^1${\it Condensed Matter Theory Group,  
The Blackett Laboratory, Imperial College, London SW7 2BZ, United Kingdom} \\
    $^2${\it Departamento de F{\'{\i}}sica Te\'orica
de la Materia Condensada, Facultad de Ciencias (C-V),
 Universidad Aut\'onoma de Madrid, 
 E-28049 Madrid, Spain} \\
}


\maketitle

\begin{abstract}
In this letter we show how transmission metallic gratings 
with very narrow and deep enough 
slits can exhibit transmission resonances 
for wavelengths larger than the period of the grating.
By using a transfer matrix formalism and a quasi-analytical 
model based on a modal expansion, we show that there are two 
possible ways of transferring light from the upper surface  
to the lower one:  
by the excitation of  
coupled surface plasmon polaritons on both surfaces  
of the metallic grating or by the coupling of incident 
plane waves with waveguide resonances located in the slits.
Both mechanisms can lead to almost perfect transmittance 
for those particular resonances.
\end{abstract}


]

\narrowtext    

Absorption anomalies in metallic 
gratings
have attracted much attention since their discovery by Wood \cite{Wood} in 1902.
One of these anomalies is
only observed for p-polarized light ({\bf H} parallel to
the grating grooves) and 
appears as a minimum on the specular reflectance. 
Now it is well known \cite{Rather} that this anomaly 
stems from the excitation of
surface plasmon polaritons
(SPPs) by the incident electromagnetic radiation. The dependence of
these SPP modes
on the grating shape and the possible existence of modes
localized in the grooves have been studied 
for a number of years and are still of interest 
\cite{Rather,Weber,Wirgin84,Lopez84,Garcia,Maradudin,Sobnack}. 
On the other hand, the activity of the last decade in the
field of photonic crystals \cite{Joannopoulos} 
has originated a renewed interest in the properties of SPPs,  
as they can be viewed as 
surface electromagnetic modes propagating in 1D 
periodic dielectric media \cite{Barnes}.

Besides, experimental evidence of the excitation of optical waveguide
modes inside the narrow grooves of zero-order 
reflection gratings has been recently given \cite{Lopez}.
Both from fundamental and practical points of view it would be very 
interesting to analyze the transmission properties of 
these waveguide modes. 
Also very recently some experiments 
carried out in arrays of submicrometre cylindrical 
holes in metallic films have shown an extraordinary optical  
transmission at wavelengths 
up to ten times 
larger than the diameter of the holes \cite{Ebbesen}.
The similarities between this last structure 
and metal gratings
suggest the possibility of equivalent resonant effects 
in transmission metal gratings of very narrow slits.

In this letter we test theoretically this 
possibility by analyzing the  
response of  
transmission metallic gratings  
to $p$-polarized electromagnetic radiation. 
We will show how,  for very narrow  
slits, the coupling of the incident light 
with surface electromagnetic modes of 
the grating 
can lead to almost perfect transmission 
resonances appearing at wavelengths larger than the 
period of the grating and hence much larger 
than the lateral dimensions of the slits. 

Inset of Fig. 1 shows a schematic
view of the structures under study with the definition of
the different parameters: the period of the grating ($d$),
the width ($a$) and height ($h$) of the
slits. The substrate is characterized by a 
dielectric constant, $\epsilon$.  
Advances in material technology have allowed the
production of transmission gratings with well
controlled profiles,  
which have already been used in 
different interesting applications, 
such as polarizers  
or x-ray spectrometers \cite{Schattenburg}.
From the theoretical point of 
view, there have been some studies of these structures 
in the last few years \cite{Lochbihler,Kuta}. 
However, up to our knowledge, 
transmission properties of 
very narrow slits that are periodically structured   
remain unstudied.  

In this letter we 
consider metal gratings made of gold and we 
use fixed values for the grating period ($d=3.5\mu$m)
and the width of the slits ($a=0.5\mu$m), although the 
dependence of our results on $a$ is also addressed. 
The thickness of the metallic grating ($h$) will be varied 
between $0$ and $4 \mu$m. 
The choice of these geometrical values
is motivated by the experimental findings of
waveguide resonances reported in Ref.
\cite{Lopez} for 
reflection metal gratings with the same set 
of parameters.
Nevertheless, it should be pointed out that
the effects discussed in this letter do appear for
any other range provided $a$ is very small 
in comparison to $d$ and the frequency of the incident 
light is well below the plasma frequency of the 
metal. The dielectric function of gold is described 
using the tables reported in 
Ref. \cite{Palik}.

We have analyzed the electromagnetic properties of 
these gratings by means 
of a transfer matrix
formalism \cite{Pendry}. 
Within this formalism it is possible to
calculate transmission and reflection coefficients for 
an incoming plane wave. Subsequently, the transmittance and
reflectance of the grating as well as real-space 
electromagnetic fields can be calculated.   
Fig. 1 shows  zero-order transmittance for 
normal incident radiation on metallic
gratings in vacuum as a function of 
the wavelength of the incoming plane wave. 
The grating height ($h$) is varied in these calculations 
from $0.2$ to $4 \mu$m. 
As can be seen in Fig. 1a, for deep enough gratings
($h \ge 0.6 \mu$m)
a remarkable transmission peak
appears for a wavelength slightly larger than the grating period
(in this case $3.5 \mu$m). 
This transmission peak moves to larger 
 wavelengths as the grating height increases whereas its 
linewidth is broadened. 
And, as illustrated in Fig. 1b, subsequent transmission peaks 
emerge for deeper gratings. 
The behavior of the transmittance spectrum as a function of 
the metal thickness seems to suggest that these peaks 
could be linked to the coupling of incident plane waves with 
waveguide resonances of the slits.

In order to analyze the physical origin of these transmission 
resonances,   
we have also developed an approximated modal method.
We incorporate two main simplifications to
the exact modal method reported
in \cite{Sheng}. First, as
the frequency regime we are interested in is  
below the plasma frequency of the metal, 
surface-impedance boundary conditions (SIBC) 
\cite{Lochbihler} are imposed on the metallic boundaries,
except on the vertical walls  of the slits which
are treated as perfect metal surfaces.
Second, we only consider the fundamental eigenmode in the modal expansion
of the electric and magnetic fields inside the slits, 
which is justified in the limit where 
the wavelength of light  
is much larger than the width of the slits.
The validity of these two approximations is confirmed by
the good agreement between the results calculated
by this simplified modal method
and the numerical simulations performed with the  
transfer-matrix formalism. 
 Within this single-mode approximation the two
field amplitudes inside the slits (the one associated 
with the $e^{ik_0z}$ wave and the other with the 
$e^{-ik_0z}$ one) are proportional
to $1/D$, where the denominator $D$ is given by:

\begin{eqnarray}
D = (1-(1+\eta)\phi)(1-(1+\eta)\psi)e^{\imath k_0 h} \nonumber \\
-(1+(1-\eta)\phi)(1+(1-\eta)\psi)e^{-\imath k_0 h},
\end{eqnarray}

\noindent with $k_0=2\pi/\lambda$, 
$\eta=\epsilon_{metal}^{-1/2}$, and 
$\psi$ given by the sum:

\begin{equation}
\psi = \frac{a}{d} \epsilon \sum_{m=-\infty}^{\infty}
\frac{(sinc(\frac{k_0 \gamma_m a}{2}))^2}
{(\epsilon-\gamma_m^2)^{1/2} + \epsilon \eta},
\end{equation}

\noindent where $\epsilon$ is the dielectric constant of the substrate,
$sinc(\xi) \equiv \sin(\xi)/\xi$, and 
$\gamma_m = \sin \theta + m \frac{\lambda}{d}$ is 
associated with the $m$-th diffraction order. 
The quantity $\phi$ is also given by Eq. (2) but with $\epsilon=1$. 

The zero-order transmittance spectrum 
of the  grating is completely governed by the behavior of 
the denominator $D$.  
For example, for normal incidence and gratings in vacuum,  
zero-order transmittance simplifies to:


\begin{equation}
T_0 =
\frac{16 (\frac{a}{d})^2}{\mid D \mid^2 \mid 1+\eta \mid^4}.
\end{equation}

Moreover, we have found that there is a close correspondence between 
maxima of zero-order transmittance and spectral positions of the zeros of the imaginary 
part of $D$, $\Im(D)$. 
This result allows us to analyze the 
nature of the electromagnetic modes responsible 
for the transmission resonances shown in Fig.1 just by studying 
the zeros of $\Im (D)$ as given by Eq. (1). Also, by varying 
the angle of incidence $\theta$ we can calculate the 
photonic band structure, $\omega(k_x)$, of these surface excitations. 
In Fig. 2a we show the photonic band structure for the 
case $h=0.6 \mu$m (black dots) and, for comparison, the energetic positions of 
the SPP excitation in the limit $h \rightarrow 0$ (gray dots).
Note that due to the range of photon energies we are analyzing, 
these SPP frequencies almost coincide with the 
energetic positions of the Rayleigh anomalies that are 
linked to zeros of transmittance \cite{Rather}.
As can be seen in the inset of Fig. 2a, a very narrow band gap 
between the first and second bands appears in the 
spectrum. The lower branch at $k_x=0$ 
is associated with the transmission peak 
at $\lambda$ close to $d$ in Fig. 1a. 
Its close proximity to the energy of SPP bands  
suggests that this 
 transmissive mode is associated with the 
excitation of a surface plasmon with SPP character   
in each surface of the grating. From now on, we name   
this kind of resonances as {\bf coupled SPPs}.
As $h$ is increased, new bands   
that are associated with waveguide 
modes of the slits appear in the spectrum. 
This can be seen in Fig. 2b 
which shows that for $h=3 \mu$m  
a flat waveguide band  
is present at 
$\omega=0.17$ eV. This localized mode is responsible for 
the transmission peak located at $\lambda=7.5 \mu$m  
(see Fig. 1b).   
The other transmission resonance obtained for $h=3 \mu$m 
at $\lambda \approx 5 \mu$m
corresponds to the lower branch 
of the first band gap. Differently from the $h=0.6 \mu$m case, 
a strong hybridization between SPP bands and a waveguide mode of 
similar energy results in the opening of a very broad gap. Then, the transmissive 
mode appearing at $\lambda \approx 5 \mu$m  will present a hybrid character between  
coupled SPPs and a waveguide resonance.
Therefore, by looking at the photonic band structure of 
surface plasmons, 
we can conclude that transmission resonances 
appearing in Fig. 1  
are mainly due to the excitation of two kinds of 
electromagnetic modes: coupled SPPs for  
$\lambda \approx d$ and waveguide resonances  
for $\lambda \gg d$. 
Fig. 2 can also give us additional 
information about the dependence of the transmission 
resonances on the angle of incidence, $\theta$. 
A totally different dispersion relation  
for the two kinds of transmissive modes is clearly seen in Fig. 2.   
Hence, transmittance associated with coupled SPPs will 
show a strong dependence on $\theta$ 
whereas for waveguide 
resonances transmission is almost 
independent of $\theta$.   

Using the simplified modal method, we can also 
study in detail the behavior of transmission resonances  
as a function of the width of the slits, $a$. For this purpose, 
we show in Fig. 3 zero-order transmittance curves for gratings 
in vacuum of thickness (a) 
$h=0.6 \mu$m and (b) $h=3.0 \mu$m in the 
wavelength region where transmission peaks appear. 
The width of the slits is varied between $0$ and 
$1.5 \mu$m.
For the case of transmission resonances linked to 
coupled SPPs (Fig. 3a) a minimum value of $a$ 
is needed in order to couple SPPs of each surface 
of the grating.
Above this threshold 
(whose wavelength depends on the depth of the slits and for 
$h=0.6 \mu$m is around $0.2 \mu$m), the resonance is 
extremely narrow and hence these structures could be 
used as filters of electromagnetic radiation for   
wavelengths close to the period of the grating. On the other hand, 
for waveguide resonances  
(Fig. 3b), even for extremely narrow slits 
the transmission peak could be close to 1. In this limit, 
the wavelength of the resonance tends to $2h$  
(that corresponds to the first zero of $\sin k_0 h$) and its 
linewidth goes to zero. As 
shown in Fig. 3, transmission resonances 
associated with coupled SPPs are much narrower than the ones 
linked to waveguide modes and in both cases their 
linewidths are rapidly broadened as the width of 
the slits is increased. 

Finally, two questions remain to be answered: 
how is light transmitted from one side of the 
metallic grating to the other one by these 
electromagnetic modes? and what is the difference 
in the transmission process between the two mechanisms 
mentioned above? In order to answer these questions, 
we show in Fig. 4 detailed pictures of the {\bf E}-field 
for two cases, both with $a=0.5 \mu$m: (a) $h=0.6 \mu$m and $\lambda=3.6 \mu$m 
(that corresponds to coupled SPPs) and (b) $h=3.0 \mu$m and $\lambda=7.5\mu$m 
(example of waveguide resonance). As can be seen in Fig. 4a, the normal 
incident plane wave is exciting first a SPP in the upper 
metal surface. Although metal thickness is much larger 
than the skin depth of the metal, this SPP couples 
with the corresponding SPP mode of the lower metal surface 
through a waveguide mode located in the slits. 
Then the SPP mode of the lower surface   
can match to an outgoing propagating plane wave 
of the same frequency and momentum as the incident one,  
leading to a large transmittance. Due to the nature 
of this process, these transmission resonances are very 
sensitive to the presence of a substrate in the lower 
surface: when energies of the two SPPs involved do 
not coincide, the coupling between them is less effective 
and transmittance is severely reduced. 
The transmission process associated with waveguide resonances  
is completely different to the one described above    
for coupled SPPs. As shown in Fig. 4b, 
for these electromagnetic 
modes only the metal walls of the slits play an active 
role in the process. Incident light induces 
current densities flowing parallel to the slits' 
walls, having different signs on the two 
opposite surfaces of the slits. 
Therefore, and different from coupled SPPs,  
the transmittance associated with these waveguide 
resonances is not very sensitive to the 
refraction index of the substrate. 
 
We believe that electromagnetic modes of a nature 
similar to coupled SPPs in transmission gratings are responsible 
for the extraordinary optical transmission 
reported in hole arrays \cite{Ebbesen}.
 There are several facts that support this belief.
First, the positions of the transmission peaks 
in both structures (hole arrays and transmission gratings) 
are mainly determined by the periodicity of the 
system and are almost independent 
of the diameter of the holes or
the slits width and of the particular metal used. 
Besides, transmission resonances in 
both structures 
disperse significantly with the angle of incidence. 
However, hole arrays and gratings are two different geometries and 
the correspondence between both systems must be 
established with certain caveats 
\cite{Schroter}.  
 
In conclusion, transmission properties of 
metallic gratings with very narrow slits have been
analyzed by means of a transfer matrix formalism and a 
quasi-analytical approach based on a modal expansion. 
We have shown how for deep enough gratings, resonances in the
zero-order transmission spectra appear for wavelengths 
larger than the period of the grating. For these resonances,  
zero-order transmittance could be close to 1 besides the 
fact that the wavelength of the transmitted light is much 
larger than the lateral dimension of the slits.  
Two different transmission mechanisms have been described: excitation 
of SPPs on both surfaces of the metal grating and 
coupling of the incident light with waveguide resonances of the slits. 

We are indebted to J. S\'anchez-Dehesa and T. L\'opez-R\'{\i}os 
for many helpful discussions and L. Mart\'{\i}n-Moreno and J.J. Greffet for
a critical reading of the manuscript.
We also acknowledge partial financial support from
the Acciones Integradas Program under contract  HB-1997-0032.
J.A.P. acknowledges a postdoctoral grant from the Ministerio de Educaci\'on y Cultura
of Spain.

\newpage

\begin{figure}[h]
\caption{Inset: schematic view of the lamellar
transmission metallic gratings studied in this paper (see text).
Zero-order transmittance for a normal incident plane wave
calculated by means of
the transfer matrix formalism for lamellar metal
gratings in vacuum ($d=3.5 \mu$m, $a=0.5 \mu$m) 
for different values of the grating height ($h$), 
ranging from 0.2 $\mu$m to 4.0 $\mu$m.}
\end{figure}

\begin{figure}[h]
\caption{Photonic band structure (black dots) of the surface plasmons 
responsible for the transmission resonances appearing 
at (a) $h=0.6 \mu$m and (b) $h=3 \mu$m. 
In the same figure we plot 
the energetic positions (gray dots) of SPPs in the limit $h\rightarrow 0$. 
These bands are calculated using the simplified modal method.
In the inset of this figure (a) we show 
a closed-up picture of the opening of the first band gap 
for this case.}
\end{figure}

\begin{figure}[h]
\caption{
Zero-order transmittance curves, calculated by an approximated modal
method (see text), for metallic 
gratings of period $3.5 \mu$m in vacuum and 
thickness ({\bf a}) $h=0.6 \mu$m and ({\bf b}) $h=3.0 \mu$m as 
a function of the width of the slits, $a$,  
and wavelength of the normal incident 
light. Transmittance is shown in a gray scale 
(black: transmittance between 0.9-1.0 and 
white: transmittance between 0.0-0.1).}
\end{figure}

\begin{figure}[h]
\caption{Detailed pictures of the {\bf E}-field 
over two periods of transmission metal gratings  
($d=3.5 \mu$m, $a=0.5 \mu$m) of thickness, 
(a) $h=0.6 \mu$m and (b) $h=3.0 \mu$m in vacuum. The wavelengths 
of the normal incident radiation are for 
(a) $\lambda=3.6 \mu$m and (b) $\lambda=7.5 \mu$m, that 
correspond to different transmission peaks shown in Fig. 1.
These {\bf E}-fields have been obtained with the transfer matrix formalism. 
}
\end{figure}


\vspace*{-0.5cm}

\begin{references}

\vspace*{-1.5cm}
                       
\bibitem{Wood}
R.W. Wood, Philos. Mag. {\bf 4}, 396 (1902).

\bibitem{Rather}
H. R\"ather, {\it Surface Plasmons on Smooth
and Rough Surfaces and on Gratings} 
(Springer-Verlag, Berlin, 1988).

\bibitem{Weber}
M. Weber and D.L. Mills, Phys. Rev. B {\bf 27}, 2698 (1983).

\bibitem{Wirgin84}
A. Wirgin and T. L\'opez-Rios, Opt. Commun. {\bf 48},
416 (1984); Opt. Commun. {\bf 49}, 455(E) (1984).

\bibitem{Lopez84}
T. L\'opez-Rios and A. Wirgin, Solid State Commun. {\bf 52},
197 (1984).


\bibitem{Garcia}
F.J. Garc\'{\i}a-Vidal and J.B. Pendry, Phys. Rev. Lett. 
{\bf 77}, 1163 (1996).

\bibitem{Maradudin}
A.A. Maradudin, A.V. Shchegrov, and T.A. Leskova,
Opt. Commun. {\bf 135}, 352 (1997).

\bibitem{Sobnack}
M.B. Sobnack {\it et al.}, Phys. Rev. Lett.
 {\bf 80}, 5667 (1998).


\bibitem{Joannopoulos}
J.D. Joannopoulos, R.D. Meade, and J.N. Winn, {\it Photonic Crystals}
(Princeton University Press, Princeton, 1995). 

\bibitem{Barnes} W.L. Barnes {\it et al.}, Phys. Rev. B {\bf 51}, 11164 (1995);
 W.L. Barnes {\it et al.}, Phys. Rev. B {\bf 54}, 6227 (1996).

\bibitem{Lopez}
T. L\'opez-Rios {\it et al.}, Phys. Rev. Lett. {\bf 81}, 665 (1998).

\bibitem{Ebbesen}
T.W. Ebbesen {\it et al.}, Nature (London) {\bf 391}, 667 (1998).

\bibitem{Schattenburg}
M.L. Schattenburg {\it et al.}, Opt. Eng. {\bf 30},
1590 (1991).

\bibitem{Lochbihler}
H. Lochbihler and R.A. Depine, Appl. Opt. {\bf 32},
3459 (1993);
H. Lochbihler, Phys. Rev. B {\bf 50}, 4795 (1994). 
 
\bibitem{Kuta}
J.J. Kuta {\it et al.}, J. Opt. Soc. Am. A {\bf 12}, 1118 (1995).

\bibitem{Palik}
{\it Handbook of Optical Constants of
Solids }, edited by E.D. Palik  (Academic, Orlando, 1985).
 
\bibitem{Pendry}
J.B. Pendry, J. Mod. Opt. {\bf 41}, 209 (1994);
P.M. Bell {\it et al.}, Comp. Phys. Commun. {\bf 85},
306 (1995).

\bibitem{Sheng}
P. Sheng, R.S. Stepleman, and P.N. Sanda, 
Phys. Rev. B {\bf 26},  2907  (1982).


\bibitem{Schroter}
U. Schr\"oter and D. Heitmann, Phys. Rev. B {\bf 58},
15419 (1998).

\end{references}
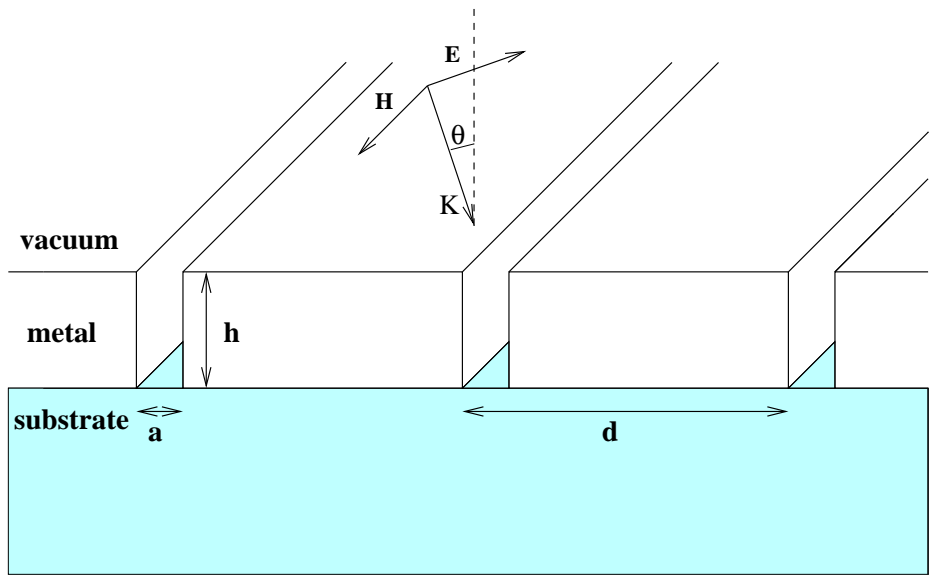
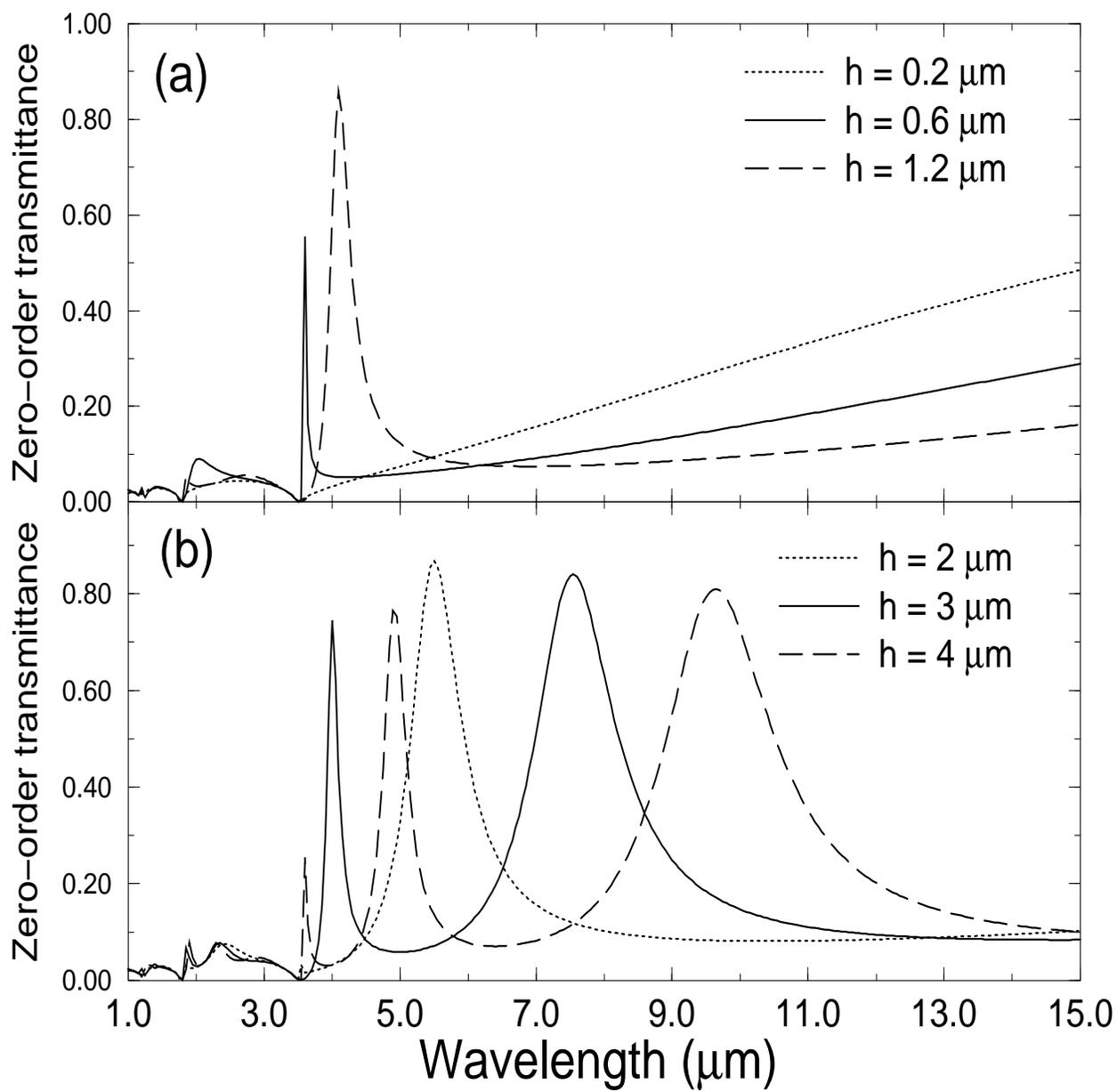
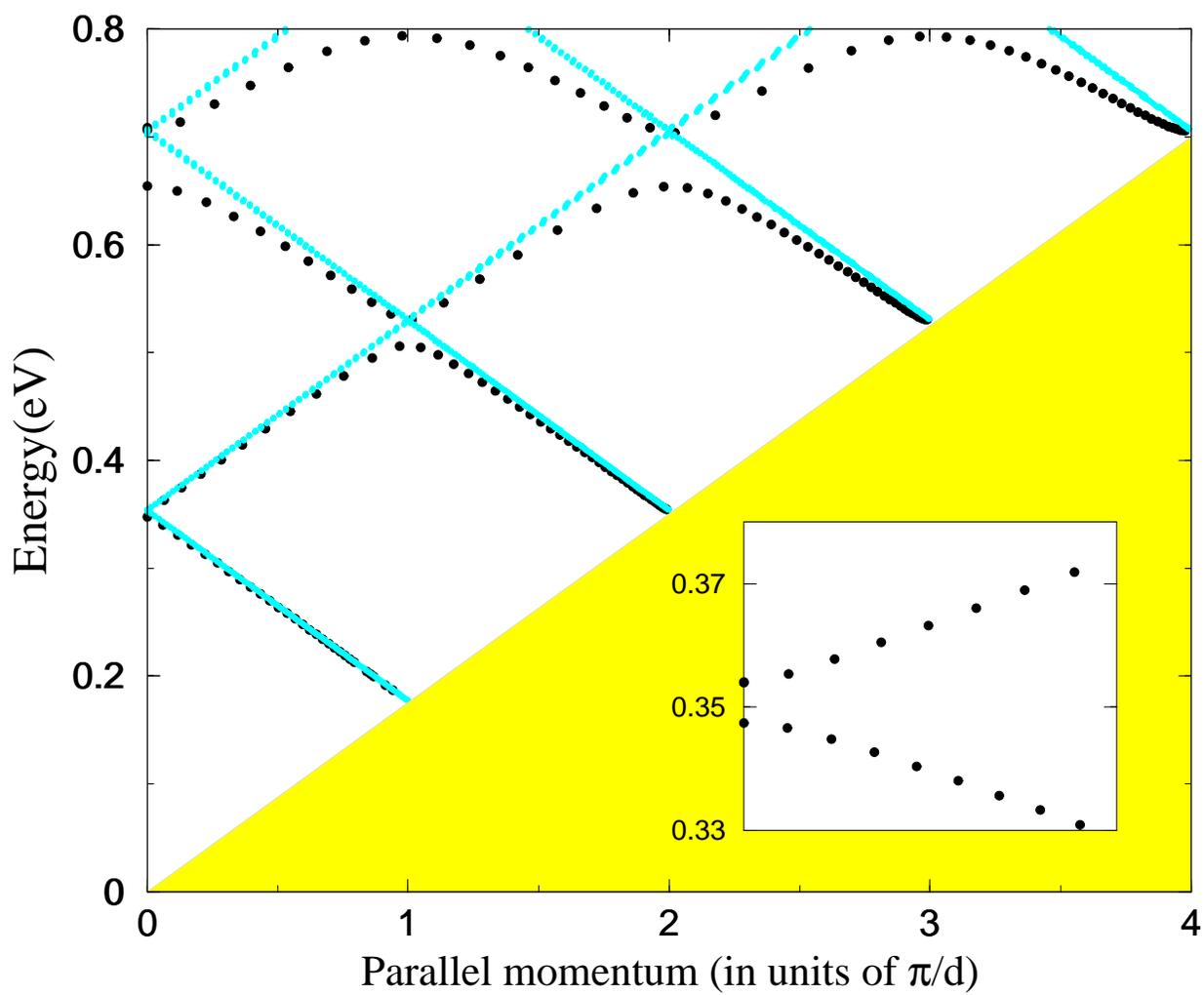
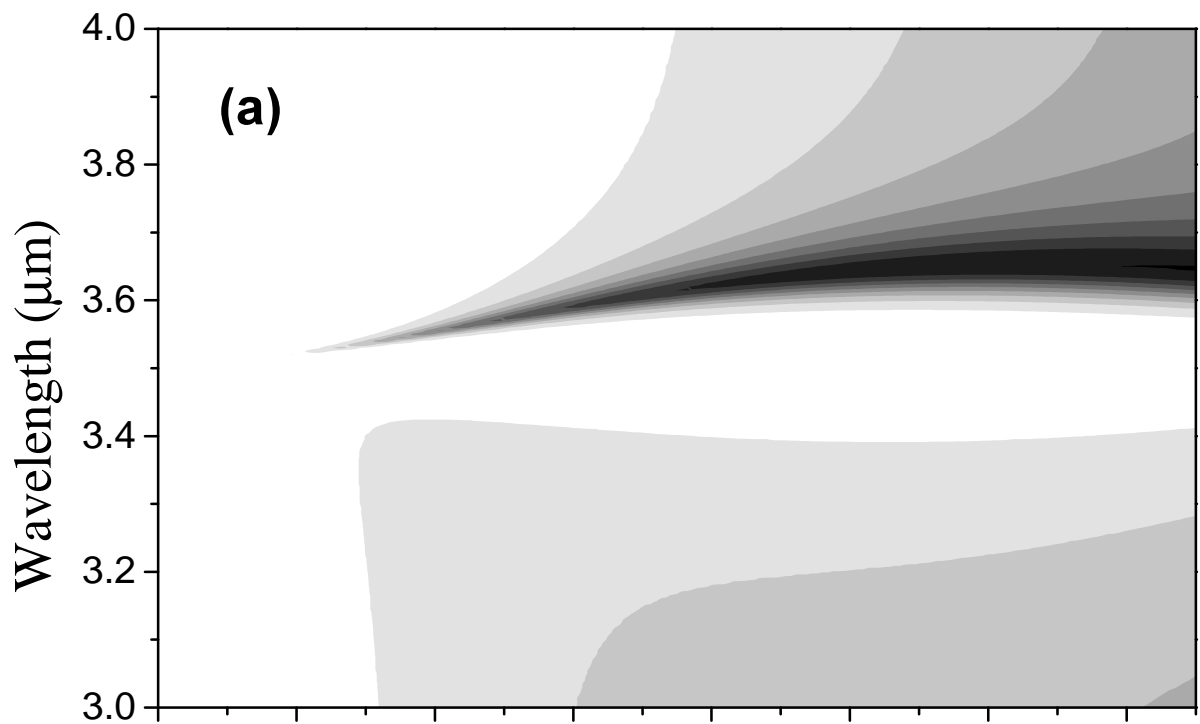
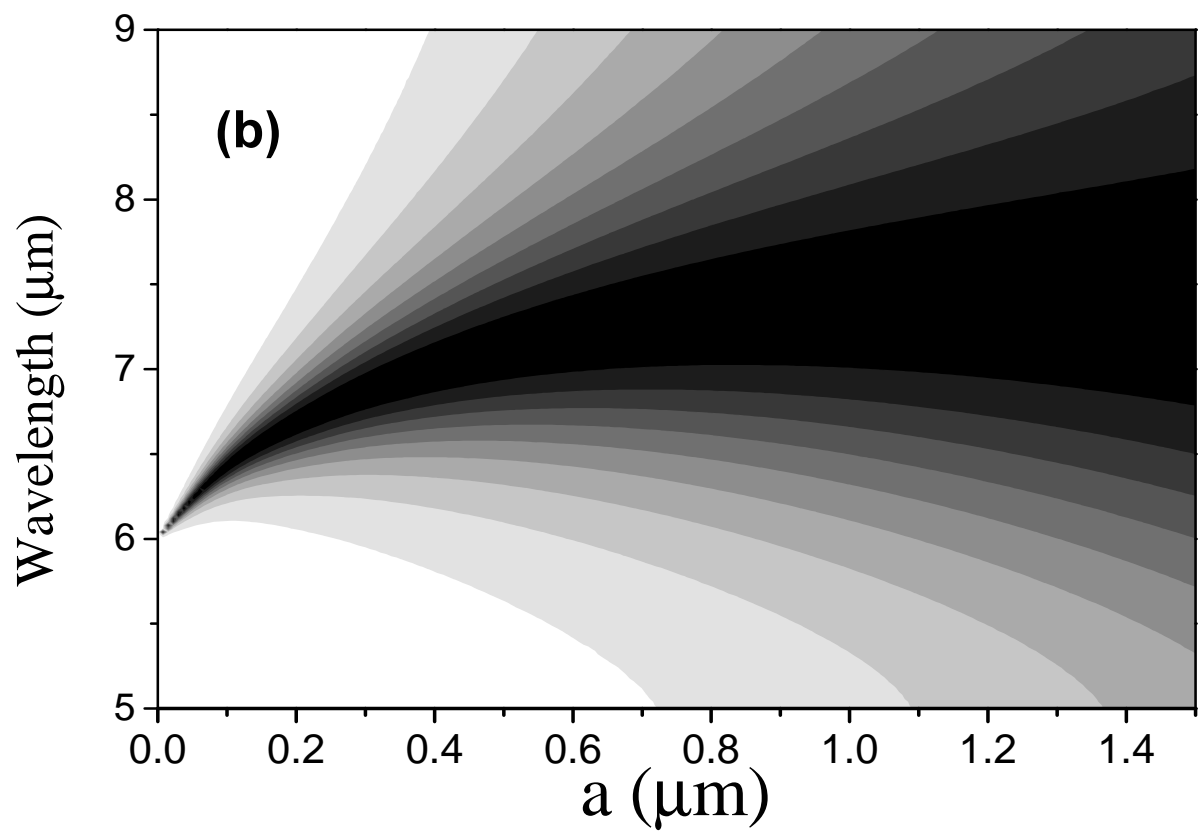
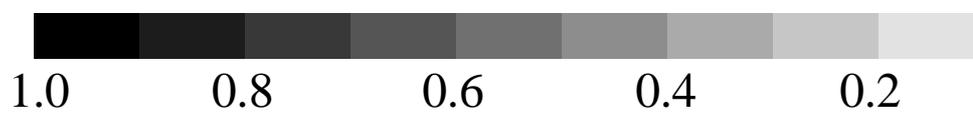
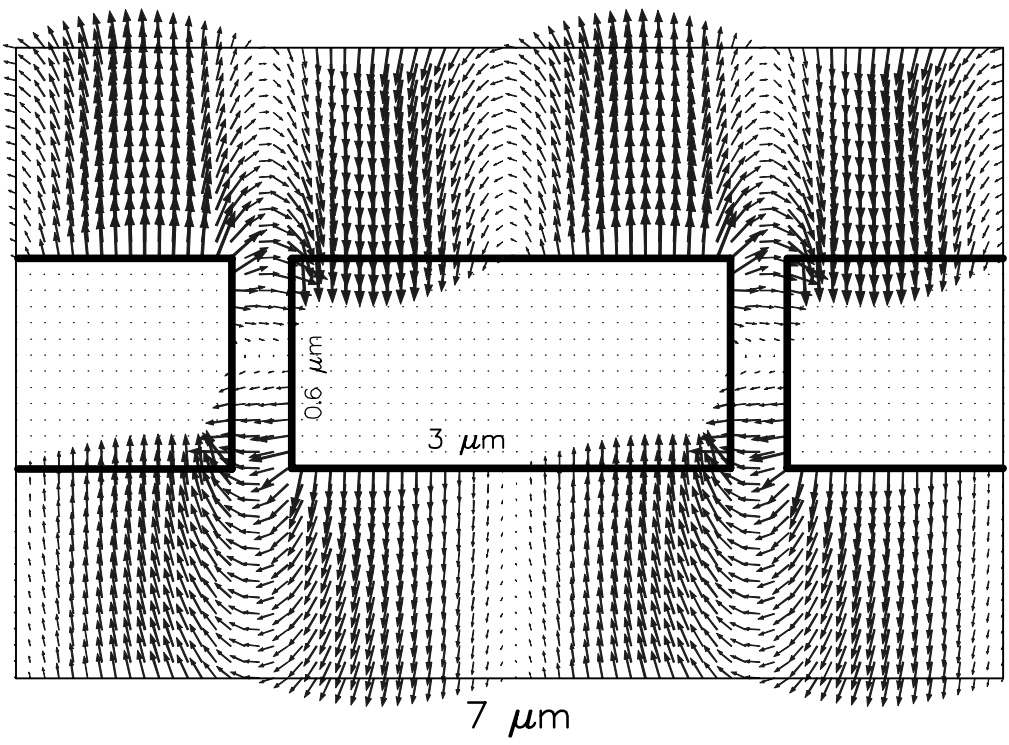
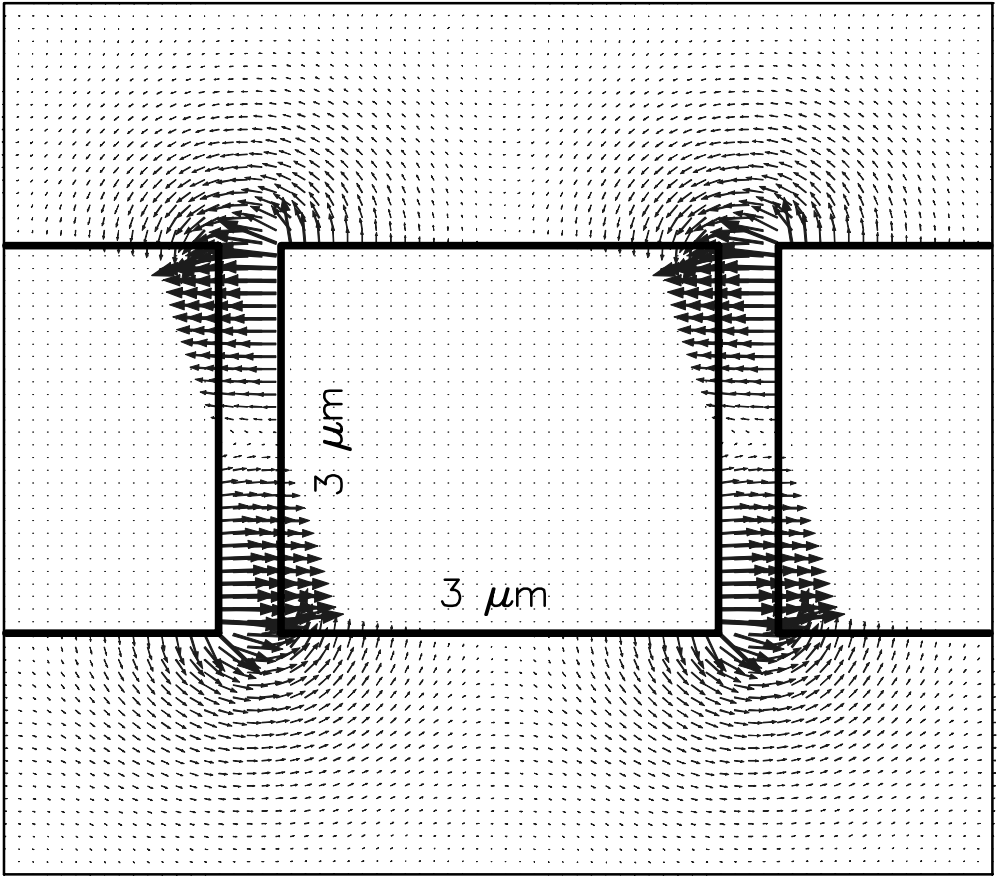
\end{document}